\documentclass[aps,prl,showpacs,twocolumn]{revtex4-1} 	

\usepackage{graphicx, amsmath, amssymb}
\usepackage{epstopdf}		

\def \erfc {\hbox{ erfc} }

\def \lvec{(\kern-.26em(}

\begin{document}

\title{Cusp Kernels for Velocity-Changing Collisions}
\author{B. H. McGuyer}
\author{R. Marsland III}
\altaffiliation{Present address: Somerville College, University of Oxford, Oxford OX2 6HD, United Kingdom.}
\author{B. A. Olsen}
\altaffiliation{Present address: Department of Physics and Astronomy and Rice Quantum Institute, Rice University, Houston, TX 77005, USA.}
\author{W. Happer}
\affiliation{Department of Physics, Princeton University, Princeton, New Jersey 08544, USA}
\date{\today}%

\begin{abstract}
We introduce an analytical kernel, the ``cusp'' kernel, to model the effects of velocity-changing collisions on optically pumped atoms in low-pressure buffer gasses.  Like the widely used Keilson-Storer kernel [J.~Keilson and J.~E.~Storer, Q.~Appl.~Math.~{\bf 10}, 243 (1952)], cusp kernels are characterized by a single parameter and preserve a Maxwellian velocity distribution.  Cusp kernels and their superpositions are more useful than KS kernels because they are more similar to real kernels inferred from measurements or theory, and are easier to invert to find steady-state velocity distributions.
\end{abstract}


\maketitle


Collision kernels $W(x,y)$ are often used to model the effects of binary, velocity-changing collisions in experiments with dilute, optically pumped atoms in low-pressure buffer gases \cite{shalagin:book, berman:1982, rogers:1991}.
Collision kernels are connected to the transport coefficients of classical transport theory \cite{berman:1986, kryszewski:1997}.
By ``collision" we mean a statistical ensemble of collisions with impact parameters $b$ less than some  maximum value $b_m$, above which the collisional effects are negligible, and including all orbital planes---or the partial wave equivalent.
If a pumped atom has a velocity $y$ along the direction of a pumping beam before a collision, the post-collision probability to find the atom with a velocity  between $x$ and $x+dx$ along the same direction is  $dx W(x,y)$.
We express these velocities in units of the most-probable three-dimensional speed $v_D = \sqrt{2 k_B T/M}$ for atoms of mass $M$ and temperature $T$, where $k_B$ is Boltzmann's constant.
One of the most important applications of this physics is the backscattering of light by sodium atoms at an altitude of 90--100~km above the Earth in laser guidestar systems, which are extensively used in modern ground-based telescopes to compensate for atmospheric turbulence \cite{ Guidestar, holzlohner:2010}.

For many years, models of experiments have most often used the Keilson-Storer (KS) collision kernel \cite{KS}, which has the simple analytical form
\begin{align}
W_a(x,y) = \frac{ e^{-(x-ay)^2/b^2} }{ b \sqrt{\pi} }.
\label{int2}
\end{align}
The KS kernel has a single, real ``memory parameter" $a$, with $0 \leq a < 1$ and a corresponding width $b = \sqrt{1-a^2}$.
The KS kernel is normalized as a probability distribution,
\begin{align}
\int W_a(x,y) dx = 1,
\label{int4}
\end{align}
and has no effect on a Maxwellian velocity distribution, $e^{-x^2}/\sqrt{\pi}$, since
\begin{align}
\int W_a(x,y)e^{-y^2}dy = e^{-x^2}.
\label{int6}
\end{align}
Collision kernels that do not satisfy the constraint (\ref{int6}) produce non-Maxwellian steady-state velocity distributions, which are not states of maximum entropy.

KS kernels, however, are only modestly similar to real kernels measured by experiments or modeled from realistic interatomic potentials \cite{gorlicki:1982, yodh:1986, HoChu, GG, gibble:1991, kasai:2003}.
The most striking difference is that real kernels display a sharp peak near the initial velocity $y$, due to weak collisions (or small-angle scattering), that is absent in KS kernels.
More realistic model kernels with sharp peaks often lack the simplicity of KS kernels or violate the equilibrium requirement (\ref{int6}) \cite{Aminoff, zhu:1986, haverkort:1987, ocallaghan:1989}.
Here, we discuss a ``cusp'' kernel $C_s(x,y)$, a model kernel parameterized by a single ``sharpness parameter" $s$, that satisfies (\ref{int4}) and (\ref{int6}), but is more similar to real kernels than the KS kernel, and is much easier to invert to find steady-state velocity distributions.

Let $\rho(x,t) dx$ be the number of atoms at time $t$ with velocity between $x$ and $x+dx$ in a spin-polarized mode of a density matrix.
A variant of the Boltzmann equation describes the evolution of $\rho$ by
\begin{align}
\frac{\partial}{\partial t} \rho(x,t) =& - (\gamma_\text{sd} + \gamma_\text{vd})\rho(x,t) \nonumber \\
	& + \gamma_\text{vd}\int W(x,y)\rho(y,t)dy + P(x,t).
\label{int8}
\end{align}
Collisions cause the spin polarization to relax at the velocity-independent rate $\gamma_\text{sd}$.
For population spin modes, like longitudinal spin polarization, $\gamma_\text{sd}$ is a nonnegative real number.
For coherence spin modes, like transverse spin polarization, the Bohr frequency of the mode is included as an imaginary part of $\gamma_\text{sd}$.
Velocity-changing collisions transfer atoms with pre-collision velocities between $y$ and $y+dy$ to atoms with velocity $x$ at the rate $\gamma_\text{vd} W(x,y)$.  
Velocity-selected spin polarization is produced by optical pumping at a rate $P(x,t)$. For a monochromatic pumping laser one can approximate $P(x,t)\approx p\,\delta(x-x_l)$ where $x_l$ is the velocity of atoms that have been Doppler-shifted into resonance with the laser light, and $p$ parameterizes the pumping rate.

Snider \cite{Snider} has shown that the KS kernel (\ref{int2}) can be written in terms of its right and left eigenfunctions as 
\begin{align}
W_a(x,y) = \sum_{n=0}^{\infty}a^nv_n(x) v_n^L(y).
\label{int10}
\end{align}
The right and left eigenfunctions are
\begin{align}\label{int11a}
v_n(x) &= \frac{ H_n(x) e^{-x^2} }{ \sqrt{ 2^n n! \pi} } \\
\text{and} \quad v_n^L(y)&= \frac{ H_n(y) }{ \sqrt{ 2^n n! } },
\label{int11}
\end{align}
respectively, where $H_n(x)$ is a Hermite polynomial:  $H_0(x) = 1, H_1(x) = 2x, H_2(x) = 4x^2 - 2, \ldots$
The right and left eigenfunctions are orthonormal, $\int v_n^L(x)v_m(x) dx = \delta_{nm}$, and complete, $\sum_n v_n(x) v_n^L(y) = \delta(x-y)$.

\begin{figure}[t!]
	\centering
	\includegraphics{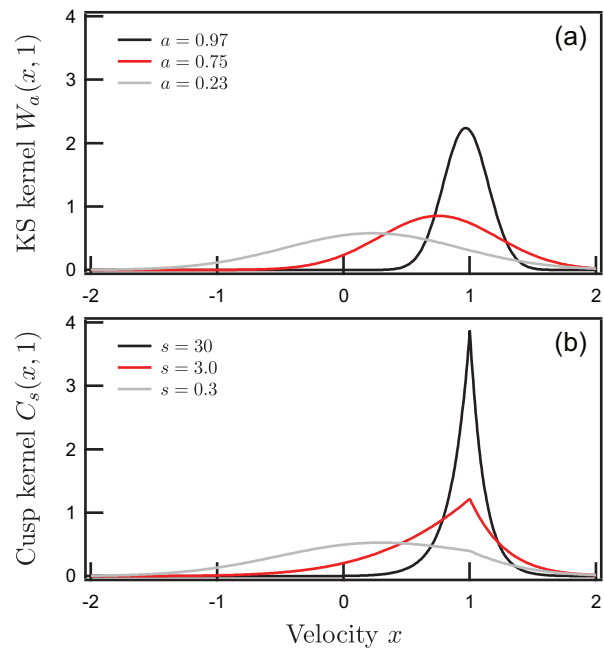}
	\caption{\label{Fig1}(color online) A comparison of KS kernels (top) with various memory parameters $a$ and cusp kernels (bottom) with various sharpness parameters $s$. In contrast to KS kernels, cusp kernels display a sharp peak near the initial velocity $y = 1$.}
\end{figure}

In reality, there are many more ``weak'' (or grazing-incidence) collisions, corresponding to a KS memory
parameter $a\approx 1$, than ``strong'' (or head-on) collisions with  $a \approx 0$.
Simple KS kernels do not capture this dominance of weak collisions.
A better model would be a superposition of KS kernels  weighted to have the most probable value of $a$ close to 1.
Such a probability density is
\begin{equation}
P_s(a) = s a^{s-1},
\label{int12}
\end{equation}
where we will call the nonzero parameter $s$ the ``sharpness.''
For large positive sharpnesses, $s \gg 1$, the probability density (\ref{int12}) heavily weights memory parameters $a \approx 1$.
We call the kernel produced by the probability density (\ref{int12}) and (\ref{int10}),
\begin{align}
C_s(x,y) &=\int_0^{1}W_a(x,y) P_s(a)da \nonumber \\ 
	&= \sum_{n=0}^{\infty} \left( \frac{s}{s+n} \right) v_n(x) v_n^L(y),
\label{int14}
\end{align}
a cusp kernel.
Cusp kernels are normalized, as in (\ref{int4}), and they transform a Maxwellian distribution into itself, as in (\ref{int6}).
Representative KS and cusp kernels are shown in Fig.~\ref{Fig1}.
Cusp kernels often resemble hard-sphere kernels  \cite{kolchenko:1972, liao:1980, belai:2007}.
For $s > 1$, cusp kernels display a sharp peak near the initial velocity $y$, with a sharpness that increases with the value of the parameter $s$.

Morgan and Happer \cite{MH} have shown that (\ref{int14}) can be written as
\begin{align}
C_s(x,y) = \frac{s 2^s \Gamma(s)}{\sqrt{\pi}} e^{y^2} R_s(-x_<)R_s(x_>),
\label{int16}
\end{align}
where $\Gamma(s)$ is the Euler gamma function, $x_>$ is the greater of the two arguments $x$ and $y$, and $x_<$ is the lesser.
For integer values of the sharpness $s$, the right functions $R_s(z)$ can be written as
\begin{align}
R_n(z)=\left \{\begin{array}{ll}\frac{1}{2}\sqrt{\pi}\erfc_{\,n-1}(z),&n=0,+1,+2,\ldots \\
e^{-z^2}H_{|n|}(z), &n=0,-1,-2,\ldots\end{array}\right .
\label{int18}
\end{align}
Here, $\erfc_{\,n}(z)$ denotes a repeated integral of the error function, as discussed in Sec.~7.2 of Abramowitz and Stegun \cite{Abramowitz}.  For noninteger $s$, the right function $R_s(z)$ can be evaluated with the power series
\begin{align}
R_s(z)=\sum_{n=0}^{\infty}\frac{\sqrt{\pi}(-z)^n}{n!2^{s-n}\Gamma(\frac{1}{2}+\frac{s}{2}-\frac{n}{2})}.
\label{int20}
\end{align}
$R_s(z)$ is an entire function of both $z$ and $s$. Using asymptotic expressions for $R_s$ \cite{MH}, we find an approximation for $C_s(x,y)$ that gives values almost the same as those of (\ref{int16}) for  $|s|\gg 1$ and for both $|x|$ and
$|y|$ of order $1$,
\begin{align}
2\ln C_s(x,y)\simeq y^2-x^2-2|x-y|\sqrt{2s}+\ln\left(\frac{s}{2}\right).
\label{int22}
\end{align}
 
The inverse of the transformation (\ref{int14}) is
\begin{align}
W_a(x,y)=\frac{1}{2\pi i}\int_{c-i\infty}^{c+i\infty} \frac{1}{s}C_s(x,y)e^{-s\ln a} ds,
\label{int24}
\end{align}
where $c>0$ is any positive number.
To prove (\ref{int24}), one can represent $C_s(x,y)$ with the series (\ref{int14}), close the path of integration with an ``infinite semicircle" in the negative half of the complex $s$ plane (for $0<a<1$), and use Cauchy's residue theorem to recover the series (\ref{int10}).

For either KS kernels or cusp kernels and for time-independent pumping  we can write (\ref{int8}) as
\begin{align}
\frac{ \partial }{ \partial t } \rho(x,t) = - \int \Gamma(x,y)\rho(y,t)dy + P(x),
\label{int26}
\end{align}
where the damping kernel is
\begin{align}
\Gamma(x,y)=\sum_{n=0}^{\infty}\gamma_n v_n(x) v_n^L(y),
\label{int28}
\end{align}
and the eigenvalues of the damping kernel for KS and cusp kernels are
\begin{align}
\gamma_n=\left \{\begin{array}{ll}\gamma_\text{sd}+\gamma_\text{vd}(1-a^n) &\mbox{KS},\\
 \gamma_\text{sd}+\gamma_\text{vd} n/(n+s) &\mbox{cusp}.\end{array}\right .
\label{int30}
\end{align}
Most experiments actually measure the steady-state density matrix, the solution to (\ref{int26}) when $\partial\rho/\partial t =0$, which has the amplitude
\begin{align}
\rho(x) = \int \Gamma^{-1}(x,y) P(y)dy.
\label{int32}
\end{align}
Using (\ref{int28}) to evaluate the resolvent $\Gamma^{-1}(x,y)$, we find
\begin{align}
\Gamma^{-1}(x,y)=\sum_{n=0}^{\infty}\frac{v_n(x) v_n^L(y)}{\gamma_n}=\frac{1}{\gamma_{\infty}}+\frac{\gamma_\text{vd} \overline{W}(x,y)}{\gamma_{\infty}\gamma_0},
\label{int34}
\end{align}
where $\gamma_{0}=\gamma_\text{sd}$ and $\gamma_{\infty}=\gamma_\text{sd}+\gamma_\text{vd}$.
We see from (\ref{int32}) that the first term of (\ref{int34}), $1/\gamma_\infty$, which corresponds to no velocity-changing collisions, ``imprints'' the primary velocity distribution $P$ from the laser source onto the steady-state velocity distribution $\rho$.
The second term gives a collisional background (or pedestal) that is spread out over additional velocities, as described by the ``resolvent kernel'' $\overline{W}(x,y)$.
For KS and cusp collision kernels the resolvent kernels are, respectively,
\begin{align}\label{int36}
\overline{W}(x,y)&=\sum_{k=0}^{\infty}\frac{\gamma_0\gamma_\text{vd}^k}{(\gamma_{\infty})^{k+1}}W_{a^{k+1}}(x,y) \\
\text{and} \quad \overline{W}(x,y)&=C_r(x,y),\quad\hbox{where } r = s(\gamma_0/\gamma_{\infty}).
\label{int38}
\end{align}
If we represent the collision kernel with a cusp kernel $C_s(x,y)$ of sharpness $s$, then the resolvent kernel (\ref{int38}) is simply another cusp kernel $C_r(x,y)$, defined by (\ref{int14}) or (\ref{int16}), with a diminished sharpness $r = s (\gamma_0/\gamma_{\infty})$.
In contrast, if the collision kernel is represented with a KS kernel, the resolvent kernel is an infinite series (\ref{int36}) of KS kernels with the sequence of modified memory parameters $a,a^2,a^3,\ldots$, which converges rather slowly unless the parameter $a$ is nearly zero.

\begin{figure}[b!]
	\centering
	\includegraphics{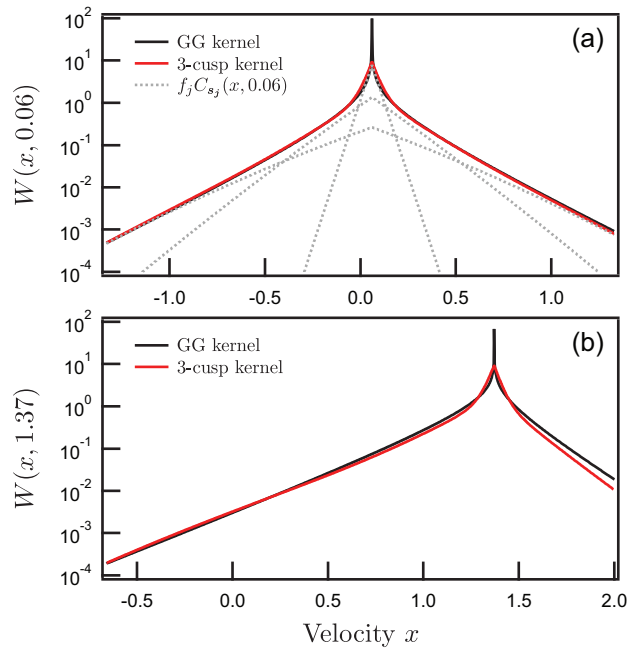}
	\caption{\label{Fig2} (color online) A comparison of a three-cusp fit and a measured collision kernel for Rb in He from Gibble and Gallagher (GG) \cite{GG} for the initial velocities $y= 0.06$ (top) and $y=1.37$ (bottom).
	The GG kernel is represented by a model form with parameters from Table I of \cite{GG}, normalized according to (\ref{int4}).
	The parameterized model form is estimated to agree with the measured kernel for Rb in He to within roughly 10\% over the range of velocities shown \cite{GG}.
	The GG kernel and three-cusp kernel are solid lines, with the GG kernel (black) slightly sharper than the three-cusp kernel (red).
	The three-cusp kernel was fit to the GG kernel for $y=0.06$ (top), which gave the superposition parameters $[f_1, f_2, f_3] = [0.13,0.37,0.50]$ and $[s_1, s_2, s_3] = [7.8,27.2,500]$, as defined by  (\ref{int39}).
	The individual cusps $f_jC_{s_j}(x,0.06)$ are shown by the dashed lines in the top panel.
	As the bottom panel shows, the same fit parameters give good agreement with measurements for $y=1.37$.
	See Fig.~10 of GG \cite{GG} for a comparison with KS fits to measurements.
	}
\end{figure}

To better approximate a real collision kernel, one could superpose $m > 1$ cusp kernels with sharpnesses $s_1, s_2,\ldots, s_m$, and with corresponding weights $f_1, f_2, \ldots, f_m$ that sum to unity, $\sum_j f_j = 1$,  as
\begin{align}\label{int39}
W(x,y) &= \sum_{j=1}^m f_j C_{s_j}(x,y).
\end{align}
The multi-cusp kernel (\ref{int39}) satisfies the constraints (\ref{int4}) and (\ref{int6}).
The second line of (\ref{int30}) is then
\begin{align}
\gamma_n = \gamma_\text{sd} + \gamma_\text{vd} \sum_{j=1}^m \frac{f_j n}{n + s_j}=\frac{N(n)}{D(n)}.
\label{int40}
\end{align}
We can write the denominator and numerator of (\ref{int40}) as the polynomials
\begin{align}
D(n) &= (n+s_1)(n+s_2)\cdots(n+s_m) \\
\text{and} \quad N(n) &= \gamma_\infty (n + r_1)(n+r_2)\cdots (n+r_m),
\label{int42}
\end{align}
where we have introduced the roots $r_j$ of $N(n)$.
Substituting (\ref{int40}) into (\ref{int34}) and expanding $D(n)/N(n)$ in partial fractions, we find that the resolvent kernel is
\begin{align}
\overline{W}(x,y) &= \sum_{j=1}^m g_j C_{r_j} (x,y).
\end{align}
Here, $C_{r_j}(x,y)$ is  a cusp kernel of sharpness $r_j$, and the weight coefficients $g_j$, which sum to unity, $\sum_j g_j = 1$, are
\begin{align}
g_j = \lim_{n\to\,-r_j}\frac{\gamma_\infty \gamma_0(n+r_j)D(-r_j)}{\gamma_\text{vd} r_j N(n) }.
\end{align}
We see that if the damping kernel $\Gamma(x,y)$ contains $m$ cusp kernels of sharpnesses $s_1, s_2,\ldots, s_m$, the resolvent $\Gamma^{-1}(x,y)$ also contains $m$ cusp kernels with modified sharpnesses $r_1, r_2,\ldots, r_m$ given by (\ref{int42}).
There does not seem to be an analogous, simple relation for the resolvent of a damping kernel with multiple KS kernels.
An example of how well a sum of cusp kernels can approximate an experimentally inferred kernel \cite{GG} is shown in Fig.~2.
As the figure demonstrates, cusp kernels are a convenient basis to parameterize experimentally inferred kernels.

In summary, we have introduced a convenient model collision kernel, the cusp kernel $C_s(x,y)$, which is characterized by a single sharpness parameter $s$.  Like KS kernels,  cusp kernels and their superpositions are normalized, as in (\ref{int4}), and  they have a Maxwellian velocity distribution for their equilibrium state, as in  (\ref{int6}).
Compared to KS kernels, cusp kernels and their superpositions are more similar to real kernels and can be more conveniently inverted to model the steady-state velocity distributions of optically pumped atoms in low-pressure buffer gases.

\begin{acknowledgments}
The authors are grateful to Paul R.~Berman, Dima Budker, Kurt E.~Gibble, and A.~M.~Shalagin for helpful discussion.
This work was supported by the Air Force Office of Scientific Research.
\end{acknowledgments}


\begin{thebibliography}{24}%
\makeatletter
\providecommand \@ifxundefined [1]{%
 \@ifx{#1\undefined}
}%
\providecommand \@ifnum [1]{%
 \ifnum #1\expandafter \@firstoftwo
 \else \expandafter \@secondoftwo
 \fi
}%
\providecommand \@ifx [1]{%
 \ifx #1\expandafter \@firstoftwo
 \else \expandafter \@secondoftwo
 \fi
}%
\providecommand \natexlab [1]{#1}%
\providecommand \enquote  [1]{``#1''}%
\providecommand \bibnamefont  [1]{#1}%
\providecommand \bibfnamefont [1]{#1}%
\providecommand \citenamefont [1]{#1}%
\providecommand \href@noop [0]{\@secondoftwo}%
\providecommand \href [0]{\begingroup \@sanitize@url \@href}%
\providecommand \@href[1]{\@@startlink{#1}\@@href}%
\providecommand \@@href[1]{\endgroup#1\@@endlink}%
\providecommand \@sanitize@url [0]{\catcode `\\12\catcode `\$12\catcode
  `\&12\catcode `\#12\catcode `\^12\catcode `\_12\catcode `\%12\relax}%
\providecommand \@@startlink[1]{}%
\providecommand \@@endlink[0]{}%
\providecommand \url  [0]{\begingroup\@sanitize@url \@url }%
\providecommand \@url [1]{\endgroup\@href {#1}{\urlprefix }}%
\providecommand \urlprefix  [0]{URL }%
\providecommand \Eprint [0]{\href }%
\providecommand \doibase [0]{http://dx.doi.org/}%
\providecommand \selectlanguage [0]{\@gobble}%
\providecommand \bibinfo  [0]{\@secondoftwo}%
\providecommand \bibfield  [0]{\@secondoftwo}%
\providecommand \translation [1]{[#1]}%
\providecommand \BibitemOpen [0]{}%
\providecommand \bibitemStop [0]{}%
\providecommand \bibitemNoStop [0]{.\EOS\space}%
\providecommand \EOS [0]{\spacefactor3000\relax}%
\providecommand \BibitemShut  [1]{\csname bibitem#1\endcsname}%
\let\auto@bib@innerbib\@empty
\bibitem [{\citenamefont {Rautian}\ and\ \citenamefont
  {Shalagin}(1991)}]{shalagin:book}%
  \BibitemOpen
  \bibfield  {author} {\bibinfo {author} {\bibfnamefont {S.~G.}\ \bibnamefont
  {Rautian}}\ and\ \bibinfo {author} {\bibfnamefont {A.~M.}\ \bibnamefont
  {Shalagin}},\ }\href@noop {} {\emph {\bibinfo {title} {Kinetic Problems of
  Non-Linear Spectroscopy}}}\ (\bibinfo  {publisher} {North-Holland},\ \bibinfo
  {address} {Amsterdam},\ \bibinfo {year} {1991})\BibitemShut {NoStop}%
\bibitem [{\citenamefont {Berman}\ \emph {et~al.}(1982)\citenamefont {Berman},
  \citenamefont {Mossberg},\ and\ \citenamefont {Hartmann}}]{berman:1982}%
  \BibitemOpen
  \bibfield  {author} {\bibinfo {author} {\bibfnamefont {P.~R.}\ \bibnamefont
  {Berman}}, \bibinfo {author} {\bibfnamefont {T.~W.}\ \bibnamefont
  {Mossberg}}, \ and\ \bibinfo {author} {\bibfnamefont {S.~R.}\ \bibnamefont
  {Hartmann}},\ }\href {\doibase 10.1103/PhysRevA.25.2550} {\bibfield
  {journal} {\bibinfo  {journal} {Phys. Rev. A}\ }\textbf {\bibinfo {volume}
  {25}},\ \bibinfo {pages} {2550} (\bibinfo {year} {1982})}\BibitemShut
  {NoStop}%
\bibitem [{\citenamefont {Rogers}\ and\ \citenamefont
  {Berman}(1991)}]{rogers:1991}%
  \BibitemOpen
  \bibfield  {author} {\bibinfo {author} {\bibfnamefont {G.~L.}\ \bibnamefont
  {Rogers}}\ and\ \bibinfo {author} {\bibfnamefont {P.~R.}\ \bibnamefont
  {Berman}},\ }\href {\doibase 10.1103/PhysRevA.44.417} {\bibfield  {journal}
  {\bibinfo  {journal} {Phys. Rev. A}\ }\textbf {\bibinfo {volume} {44}},\
  \bibinfo {pages} {417} (\bibinfo {year} {1991})}\BibitemShut {NoStop}%
\bibitem [{\citenamefont {Berman}\ \emph {et~al.}(1986)\citenamefont {Berman},
  \citenamefont {Haverkort},\ and\ \citenamefont {Woerdman}}]{berman:1986}%
  \BibitemOpen
  \bibfield  {author} {\bibinfo {author} {\bibfnamefont {P.~R.}\ \bibnamefont
  {Berman}}, \bibinfo {author} {\bibfnamefont {J.~E.~M.}\ \bibnamefont
  {Haverkort}}, \ and\ \bibinfo {author} {\bibfnamefont {J.~P.}\ \bibnamefont
  {Woerdman}},\ }\href {\doibase 10.1103/PhysRevA.34.4647} {\bibfield
  {journal} {\bibinfo  {journal} {Phys. Rev. A}\ }\textbf {\bibinfo {volume}
  {34}},\ \bibinfo {pages} {4647} (\bibinfo {year} {1986})}\BibitemShut
  {NoStop}%
\bibitem [{\citenamefont {Kryszewski}\ and\ \citenamefont
  {Gondek}(1997)}]{kryszewski:1997}%
  \BibitemOpen
  \bibfield  {author} {\bibinfo {author} {\bibfnamefont {S.}~\bibnamefont
  {Kryszewski}}\ and\ \bibinfo {author} {\bibfnamefont {J.}~\bibnamefont
  {Gondek}},\ }\href {\doibase 10.1103/PhysRevA.56.3923} {\bibfield  {journal}
  {\bibinfo  {journal} {Phys. Rev. A}\ }\textbf {\bibinfo {volume} {56}},\
  \bibinfo {pages} {3923} (\bibinfo {year} {1997})}\BibitemShut {NoStop}%
\bibitem [{\citenamefont {Happer}\ \emph {et~al.}(1994)\citenamefont {Happer},
  \citenamefont {MacDonald}, \citenamefont {Max},\ and\ \citenamefont
  {Dyson}}]{Guidestar}%
  \BibitemOpen
  \bibfield  {author} {\bibinfo {author} {\bibfnamefont {W.}~\bibnamefont
  {Happer}}, \bibinfo {author} {\bibfnamefont {G.~J.}\ \bibnamefont
  {MacDonald}}, \bibinfo {author} {\bibfnamefont {C.~E.}\ \bibnamefont {Max}},
  \ and\ \bibinfo {author} {\bibfnamefont {F.~J.}\ \bibnamefont {Dyson}},\
  }\href {\doibase 10.1364/JOSAA.11.000263} {\bibfield  {journal} {\bibinfo
  {journal} {J. Opt. Soc. Am. A}\ }\textbf {\bibinfo {volume} {11}},\ \bibinfo
  {pages} {263} (\bibinfo {year} {1994})}\BibitemShut {NoStop}%
\bibitem [{\citenamefont {Holzl\"ohner}\ \emph {et~al.}(2010)\citenamefont
  {Holzl\"ohner}, \citenamefont {Rochester}, \citenamefont {{Bonaccini Calia}},
  \citenamefont {Budker}, \citenamefont {Higbie},\ and\ \citenamefont
  {Hackenberg}}]{holzlohner:2010}%
  \BibitemOpen
  \bibfield  {author} {\bibinfo {author} {\bibfnamefont {R.}~\bibnamefont
  {Holzl\"ohner}}, \bibinfo {author} {\bibfnamefont {S.~M.}\ \bibnamefont
  {Rochester}}, \bibinfo {author} {\bibfnamefont {D.}~\bibnamefont {{Bonaccini
  Calia}}}, \bibinfo {author} {\bibfnamefont {D.}~\bibnamefont {Budker}},
  \bibinfo {author} {\bibfnamefont {J.~M.}\ \bibnamefont {Higbie}}, \ and\
  \bibinfo {author} {\bibfnamefont {W.}~\bibnamefont {Hackenberg}},\ }\href
  {\doibase 10.1051/0004-6361/200913108} {\bibfield  {journal} {\bibinfo
  {journal} {Astron. Astrophys.}\ }\textbf {\bibinfo {volume} {510}},\ \bibinfo
  {pages} {A20} (\bibinfo {year} {2010})}\BibitemShut {NoStop}%
\bibitem [{\citenamefont {Keilson}\ and\ \citenamefont {Storer}(1952)}]{KS}%
  \BibitemOpen
  \bibfield  {author} {\bibinfo {author} {\bibfnamefont {J.}~\bibnamefont
  {Keilson}}\ and\ \bibinfo {author} {\bibfnamefont {J.~E.}\ \bibnamefont
  {Storer}},\ }\href@noop {} {\bibfield  {journal} {\bibinfo  {journal} {Q.
  Appl. Math.}\ }\textbf {\bibinfo {volume} {10}},\ \bibinfo {pages} {243}
  (\bibinfo {year} {1952})}\BibitemShut {NoStop}%
\bibitem [{\citenamefont {Gorlicki}\ \emph {et~al.}(1982)\citenamefont
  {Gorlicki}, \citenamefont {Lerminiaux},\ and\ \citenamefont
  {Dumont}}]{gorlicki:1982}%
  \BibitemOpen
  \bibfield  {author} {\bibinfo {author} {\bibfnamefont {M.}~\bibnamefont
  {Gorlicki}}, \bibinfo {author} {\bibfnamefont {C.}~\bibnamefont
  {Lerminiaux}}, \ and\ \bibinfo {author} {\bibfnamefont {M.}~\bibnamefont
  {Dumont}},\ }\href {\doibase 10.1103/PhysRevLett.49.1394} {\bibfield
  {journal} {\bibinfo  {journal} {Phys. Rev. Lett.}\ }\textbf {\bibinfo
  {volume} {49}},\ \bibinfo {pages} {1394} (\bibinfo {year}
  {1982})}\BibitemShut {NoStop}%
\bibitem [{\citenamefont {Yodh}\ \emph {et~al.}(1986)\citenamefont {Yodh},
  \citenamefont {Mossberg},\ and\ \citenamefont {Thomas}}]{yodh:1986}%
  \BibitemOpen
  \bibfield  {author} {\bibinfo {author} {\bibfnamefont {A.~G.}\ \bibnamefont
  {Yodh}}, \bibinfo {author} {\bibfnamefont {T.~W.}\ \bibnamefont {Mossberg}},
  \ and\ \bibinfo {author} {\bibfnamefont {J.~E.}\ \bibnamefont {Thomas}},\
  }\href {\doibase 10.1103/PhysRevA.34.5150} {\bibfield  {journal} {\bibinfo
  {journal} {Phys. Rev. A}\ }\textbf {\bibinfo {volume} {34}},\ \bibinfo
  {pages} {5150} (\bibinfo {year} {1986})}\BibitemShut {NoStop}%
\bibitem [{\citenamefont {{T.-S.~Ho and S.-I. Chu}}(1986)}]{HoChu}%
  \BibitemOpen
  \bibfield  {author} {\bibinfo {author} {\bibnamefont {{T.-S.~Ho and S.-I.
  Chu}}},\ }\href {\doibase 10.1103/PhysRevA.33.3067} {\bibfield  {journal}
  {\bibinfo  {journal} {Phys. Rev. A}\ }\textbf {\bibinfo {volume} {33}},\
  \bibinfo {pages} {3067} (\bibinfo {year} {1986})}\BibitemShut {NoStop}%
\bibitem [{\citenamefont {Gibble}\ and\ \citenamefont {Gallagher}(1991)}]{GG}%
  \BibitemOpen
  \bibfield  {author} {\bibinfo {author} {\bibfnamefont {K.~E.}\ \bibnamefont
  {Gibble}}\ and\ \bibinfo {author} {\bibfnamefont {A.}~\bibnamefont
  {Gallagher}},\ }\href {\doibase 10.1103/PhysRevA.43.1366} {\bibfield
  {journal} {\bibinfo  {journal} {Phys. Rev. A}\ }\textbf {\bibinfo {volume}
  {43}},\ \bibinfo {pages} {1366} (\bibinfo {year} {1991})}\BibitemShut
  {NoStop}%
\bibitem [{\citenamefont {Gibble}\ and\ \citenamefont
  {Cooper}(1991)}]{gibble:1991}%
  \BibitemOpen
  \bibfield  {author} {\bibinfo {author} {\bibfnamefont {K.~E.}\ \bibnamefont
  {Gibble}}\ and\ \bibinfo {author} {\bibfnamefont {J.}~\bibnamefont
  {Cooper}},\ }\href {\doibase 10.1103/PhysRevA.44.R5335} {\bibfield  {journal}
  {\bibinfo  {journal} {Phys. Rev. A}\ }\textbf {\bibinfo {volume} {44}},\
  \bibinfo {pages} {R5335} (\bibinfo {year} {1991})}\BibitemShut {NoStop}%
\bibitem [{\citenamefont {Kasai}\ \emph {et~al.}(2003)\citenamefont {Kasai},
  \citenamefont {Mizutani}, \citenamefont {Kondo}, \citenamefont {Hasuo},\ and\
  \citenamefont {Fujimoto}}]{kasai:2003}%
  \BibitemOpen
  \bibfield  {author} {\bibinfo {author} {\bibfnamefont {S.}~\bibnamefont
  {Kasai}}, \bibinfo {author} {\bibfnamefont {R.}~\bibnamefont {Mizutani}},
  \bibinfo {author} {\bibfnamefont {R.}~\bibnamefont {Kondo}}, \bibinfo
  {author} {\bibfnamefont {M.}~\bibnamefont {Hasuo}}, \ and\ \bibinfo {author}
  {\bibfnamefont {T.}~\bibnamefont {Fujimoto}},\ }\href {\doibase
  10.1143/JPSJ.72.1936} {\bibfield  {journal} {\bibinfo  {journal} {J. Phys.
  Soc. Jpn.}\ }\textbf {\bibinfo {volume} {72}},\ \bibinfo {pages} {1936}
  (\bibinfo {year} {2003})}\BibitemShut {NoStop}%
\bibitem [{\citenamefont {Aminoff}\ \emph {et~al.}(1983)\citenamefont
  {Aminoff}, \citenamefont {Javanainen},\ and\ \citenamefont
  {Kaivola}}]{Aminoff}%
  \BibitemOpen
  \bibfield  {author} {\bibinfo {author} {\bibfnamefont {C.~G.}\ \bibnamefont
  {Aminoff}}, \bibinfo {author} {\bibfnamefont {J.}~\bibnamefont {Javanainen}},
  \ and\ \bibinfo {author} {\bibfnamefont {M.}~\bibnamefont {Kaivola}},\ }\href
  {\doibase 10.1103/PhysRevA.28.722} {\bibfield  {journal} {\bibinfo  {journal}
  {Phys. Rev. A}\ }\textbf {\bibinfo {volume} {28}},\ \bibinfo {pages} {722}
  (\bibinfo {year} {1983})}\BibitemShut {NoStop}%
\bibitem [{\citenamefont {Zhu}(1986)}]{zhu:1986}%
  \BibitemOpen
  \bibfield  {author} {\bibinfo {author} {\bibfnamefont {X.}~\bibnamefont
  {Zhu}},\ }\href {\doibase 10.1103/PhysRevA.33.251} {\bibfield  {journal}
  {\bibinfo  {journal} {Phys. Rev. A}\ }\textbf {\bibinfo {volume} {33}},\
  \bibinfo {pages} {251} (\bibinfo {year} {1986})}\BibitemShut {NoStop}%
\bibitem [{\citenamefont {Haverkort}\ \emph {et~al.}(1987)\citenamefont
  {Haverkort}, \citenamefont {Woerdman},\ and\ \citenamefont
  {Berman}}]{haverkort:1987}%
  \BibitemOpen
  \bibfield  {author} {\bibinfo {author} {\bibfnamefont {J.~E.~M.}\
  \bibnamefont {Haverkort}}, \bibinfo {author} {\bibfnamefont {J.~P.}\
  \bibnamefont {Woerdman}}, \ and\ \bibinfo {author} {\bibfnamefont {P.~R.}\
  \bibnamefont {Berman}},\ }\href {\doibase 10.1103/PhysRevA.36.5251}
  {\bibfield  {journal} {\bibinfo  {journal} {Phys. Rev. A}\ }\textbf {\bibinfo
  {volume} {36}},\ \bibinfo {pages} {5251} (\bibinfo {year}
  {1987})}\BibitemShut {NoStop}%
\bibitem [{\citenamefont {O'Callaghan}\ and\ \citenamefont
  {Cooper}(1989)}]{ocallaghan:1989}%
  \BibitemOpen
  \bibfield  {author} {\bibinfo {author} {\bibfnamefont {M.~J.}\ \bibnamefont
  {O'Callaghan}}\ and\ \bibinfo {author} {\bibfnamefont {J.}~\bibnamefont
  {Cooper}},\ }\href {\doibase 10.1103/PhysRevA.39.6190} {\bibfield  {journal}
  {\bibinfo  {journal} {Phys. Rev. A}\ }\textbf {\bibinfo {volume} {39}},\
  \bibinfo {pages} {6206} (\bibinfo {year} {1989})}\BibitemShut {NoStop}%
\bibitem [{\citenamefont {Snider}(1986)}]{Snider}%
  \BibitemOpen
  \bibfield  {author} {\bibinfo {author} {\bibfnamefont {R.~F.}\ \bibnamefont
  {Snider}},\ }\href {\doibase 10.1103/PhysRevA.33.178} {\bibfield  {journal}
  {\bibinfo  {journal} {Phys. Rev. A}\ }\textbf {\bibinfo {volume} {33}},\
  \bibinfo {pages} {178} (\bibinfo {year} {1986})}\BibitemShut {NoStop}%
\bibitem [{\citenamefont {Kolchenko}\ \emph {et~al.}(1972)\citenamefont
  {Kolchenko}, \citenamefont {Rautian},\ and\ \citenamefont
  {Shalagin}}]{kolchenko:1972}%
  \BibitemOpen
  \bibfield  {author} {\bibinfo {author} {\bibfnamefont {A.~P.}\ \bibnamefont
  {Kolchenko}}, \bibinfo {author} {\bibfnamefont {S.~G.}\ \bibnamefont
  {Rautian}}, \ and\ \bibinfo {author} {\bibfnamefont {A.~M.}\ \bibnamefont
  {Shalagin}},\ }\href@noop {} {\enquote {\bibinfo {title} {{Nuclear Physics
  Institute Semiconductor Physics Internal Report No.~46}},}\ } (\bibinfo
  {year} {1972}),\ \bibinfo {note} {(unpublished)}\BibitemShut {NoStop}%
\bibitem [{\citenamefont {Liao}\ \emph {et~al.}(1980)\citenamefont {Liao},
  \citenamefont {Bjorkholm},\ and\ \citenamefont {Berman}}]{liao:1980}%
  \BibitemOpen
  \bibfield  {author} {\bibinfo {author} {\bibfnamefont {P.~F.}\ \bibnamefont
  {Liao}}, \bibinfo {author} {\bibfnamefont {J.~E.}\ \bibnamefont {Bjorkholm}},
  \ and\ \bibinfo {author} {\bibfnamefont {P.~R.}\ \bibnamefont {Berman}},\
  }\href {\doibase 10.1103/PhysRevA.21.1927} {\bibfield  {journal} {\bibinfo
  {journal} {Phys. Rev. A}\ }\textbf {\bibinfo {volume} {21}},\ \bibinfo
  {pages} {1927} (\bibinfo {year} {1980})}\BibitemShut {NoStop}%
\bibitem [{\citenamefont {Belai}\ \emph {et~al.}(2007)\citenamefont {Belai},
  \citenamefont {Schwartz},\ and\ \citenamefont {Shapiro}}]{belai:2007}%
  \BibitemOpen
  \bibfield  {author} {\bibinfo {author} {\bibfnamefont {O.~V.}\ \bibnamefont
  {Belai}}, \bibinfo {author} {\bibfnamefont {O.~Y.}\ \bibnamefont {Schwartz}},
  \ and\ \bibinfo {author} {\bibfnamefont {D.~A.}\ \bibnamefont {Shapiro}},\
  }\href {\doibase 10.1103/PhysRevA.76.012513} {\bibfield  {journal} {\bibinfo
  {journal} {Phys. Rev. A}\ }\textbf {\bibinfo {volume} {76}},\ \bibinfo
  {pages} {012513} (\bibinfo {year} {2007})}\BibitemShut {NoStop}%
\bibitem [{\citenamefont {Morgan}\ and\ \citenamefont {Happer}(2010)}]{MH}%
  \BibitemOpen
  \bibfield  {author} {\bibinfo {author} {\bibfnamefont {S.~W.}\ \bibnamefont
  {Morgan}}\ and\ \bibinfo {author} {\bibfnamefont {W.}~\bibnamefont
  {Happer}},\ }\href {\doibase 10.1103/PhysRevA.81.042703} {\bibfield
  {journal} {\bibinfo  {journal} {Phys. Rev. A}\ }\textbf {\bibinfo {volume}
  {81}},\ \bibinfo {pages} {042703} (\bibinfo {year} {2010})}\BibitemShut
  {NoStop}%
\bibitem [{\citenamefont {Abramowitz}\ and\ \citenamefont
  {Stegun}(1965)}]{Abramowitz}%
  \BibitemOpen
  \bibinfo {editor} {\bibfnamefont {M.}~\bibnamefont {Abramowitz}}\ and\
  \bibinfo {editor} {\bibfnamefont {I.~A.}\ \bibnamefont {Stegun}},\ eds.,\
  \href@noop {} {\emph {\bibinfo {title} {Handbook of Mathematical
  Functions}}}\ (\bibinfo  {publisher} {Dover Publications},\ \bibinfo
  {address} {New York},\ \bibinfo {year} {1965})\BibitemShut {NoStop}%
\end{thebibliography}

%

%

\end{document}